\documentclass[fleqn,usenatbib]{mnras}
\usepackage{amsmath}
\usepackage{xspace}
\usepackage{mathptmx}
\usepackage{txfonts}
\usepackage{xcolor}
\usepackage{cancel}
\usepackage[T1]{fontenc}
\usepackage{ae,aecompl}

\usepackage{graphicx}	% Including figure files
\usepackage{amssymb}	% Extra maths symbols
\usepackage{booktabs}
\usepackage{wasysym}
\usepackage{float}
\usepackage{caption}

\let\oldhref\href
\renewcommand{\href}[2]{\oldhref{#1}{\hbox{#2}}}

\definecolor{colorl1}{RGB}{0, 51, 153}
\definecolor{colorl2}{RGB}{153, 0, 0}
\definecolor{colorl3}{RGB}{179, 179, 0}
\definecolor{colorl4}{RGB}{51, 102, 0}

\definecolor{colorw1}{RGB}{51, 102, 255}
\definecolor{colorw2}{RGB}{255, 51, 0}
\definecolor{colorw3}{RGB}{255, 214, 51}
\definecolor{colorw4}{RGB}{51, 204, 51}

\newcommand{\hMpc}{{\ifmmode{h^{-1}{\rm Mpc}}\else{$h^{-1}$Mpc}\fi}}
\newcommand{\Mpc}{{\ifmmode{{\rm Mpc}}\else{Mpc}\fi}}
\newcommand{\hkpc}{{\ifmmode{h^{-1}{\rm kpc}}\else{$h^{-1}$kpc}\fi}}
\newcommand{\kpc}{{\ifmmode{ {\rm kpc}}\else{{\rm kpc}}\fi}}
\newcommand{\kms}{{\ifmmode{ {\rm km\,s^{-1}}}\else{ ${\rm km\,s^{-1}}$}\fi}}
\newcommand{\hMsun}{{\ifmmode{h^{-1}{\rm {M_{\astrosun}}}}\else{$h^{-1}{\rm{M_{\astrosun}}}$}\fi}}
\newcommand{\Msun}{{\ifmmode{{\rm M}_{\astrosun}}\else{${\rm M}_{\astrosun}$}\fi}}
                        
\newcommand{\Mhalo}{{\ifmmode{M_{\rm halo}}\else{$M_{\rm halo}$}\fi}}
\newcommand{\Rvir}{{\ifmmode{R_{\rm vir}}\else{$R_{\rm vir}$}\fi}}
\newcommand{\Mvir}{{\ifmmode{M_{\rm vir}}\else{$M_{\rm vir}$}\fi}}
\newcommand{\Mstar}{{\ifmmode{M_{\rm star}}\else{$M_{\rm star}$}\fi}}
\newcommand{\Vrot}{{\ifmmode{V_{\rm rot}}\else{$V_{\rm rot}$}\fi}}
\newcommand{\ltsima}{$\; \buildrel < \over \sim \;$}
\newcommand{\gtsima}{$\; \buildrel > \over \sim \;$}
\newcommand{\lsim}{\lower.5ex\hbox{\ltsima}}
\newcommand{\gsim}{\lower.5ex\hbox{\gtsima}}

\def\lesssim{\mathrel{\hbox{\rlap{\hbox{\lower4pt\hbox{$\sim$}}}\hbox{$<$}}}}
\def\gtrsim{\mathrel{\hbox{\rlap{\hbox{\lower4pt\hbox{$\sim$}}}\hbox{$>$}}}}

\newcommand{\beq}{\begin{equation}}
\newcommand{\eeq}{\end{equation}}
\def\beqa{\begin{eqnarray}}
\def\eeqa{\end{eqnarray}}
\def\LCDM{\ensuremath{\Lambda}CDM}

\def\head{ \vbox to 0pt{\vss \hbox to 0pt{\hskip 440pt\rm
      LA-UR-10-07069\hss} \vskip 25pt}}

\def \kms {\ifmmode  \,\rm km\,s^{-1}\else $\,\rm km\,s^{-1}$\fi }
\def \kpc {\ifmmode  {\,\rm kpc}  \else ${\rm  kpc}$ \fi  }  
\def \hkpc {\ifmmode  {h^{-1}\rm kpc}  \else ${h^{-1}\rm kpc}$ \fi  }  
\def \hMpc {\ifmmode  {h^{-1}\rm Mpc}  \else ${h^{-1}\rm Mpc}$ \fi  }  
\def \Mpch {\ifmmode  {h^{-1}\rm Mpc}  \else ${h^{-1}\rm Mpc}$ \fi  }  
\def \Msun {\ifmmode {\rm M}_{\astrosun} \else ${\rm M}_{\astrosun}$ \fi} 
\def \hMsun {\ifmmode h^{-1}\,\rm M_{\astrosun} \else $h^{-1}\,\rm M_{\astrosun}$ \fi}
\def \Gyr {\ifmmode\, \rm Gyr \else $\,$Gyr \fi}

\def \LCDM {\ifmmode \Lambda{\rm CDM} \else $\Lambda{\rm CDM}$ \fi}
\def \sig8 {\ifmmode \sigma_8 \else $\sigma_8$ \fi} 
\def \OmegaM {\ifmmode \Omega_{\rm m} \else $\Omega_{\rm m}$ \fi} 
\def \Omegab {\ifmmode \Omega_{\rm b} \else $\Omega_{\rm b}$ \fi} 
\def \OmegaL {\ifmmode \Omega_{\rm \Lambda} \else $\Omega_{\rm \Lambda}$\fi} 
\def \Deltavir {\ifmmode \Delta_{\rm vir} \else $\Delta_{\rm vir}$ \fi}
\def \rhocrit {\ifmmode \rho_{\rm crit} \else $\rho_{\rm crit}$ \fi}
\def \rhou {\ifmmode \rho_{\rm u} \else $\rho_{\rm u}$ \fi}
\def \zc {\ifmmode z_{\rm c} \else $z_{\rm c}$ \fi}

\def\Mstar {\ensuremath {M_{*}(<r_{23.5})}~}
\def\r23_5 {\ensuremath {r_{23.5}}~}

\title[AI to constrain simulations] {Using Artificial Intelligence and real galaxy images to constrain parameters in galaxy formation simulations}

\author[A.V. Macci\`o et al.]{Andrea V. Macci\`o$^{1,2,3}$\thanks{E-mail: maccio@nyu.edu},
Mohamad Ali-Dib$^{2,1}$, Pavle Vulanović$^{1,2}$, Hind Al Noori$^1$, 
\newauthor{Fabian Walter$^{3}$, Nico Krieger$^3$,  Tobias Buck$^4$}\\
$^{1}$New York University Abu Dhabi, PO Box 129188, Abu Dhabi, United Arab Emirates \\
$^2$Center for Astro, Particle and Planetary Physics (CAP$^3$), New York University Abu Dhabi\\
$^3$Max-Planck-Institut f\"ur Astronomie, K\"onigstuhl 17, 69117 Heidelberg, Germany \\
$^4$Leibniz-Institut f\"ur Astrophysik Potsdam (AIP), An der Sternwarte 16, D-14482 Potsdam, Germany\\
}

\date{Accepted XXX. Received YYY; in original form ZZZ}

\pubyear{2021}

\begin{document}

\label{firstpage}
\pagerange{\pageref{firstpage}--\pageref{lastpage}}
\maketitle
\begin{abstract}

Cosmological galaxy formation simulations are still limited by their spatial/mass resolution and cannot model from first principles some of the processes, like star formation, that are key in driving galaxy evolution. As a consequence they still rely on a set of 'effective parameters' that try to capture the scales and the physical processes that cannot be directly resolved in the simulation. In this study we show that it is possible to use Machine Learning techniques applied to real and simulated images of galaxies to discriminate between different values of these parameters by making use of the full {information content} of an astronomical image instead of collapsing it into a limited set of values like  size, or stellar/ gas masses. 
{In this work} we apply our method to the NIHAO simulations and the THINGS and VLA-ANGST observations of HI maps in nearby galaxies to test the ability of different values of the star formation density threshold $n$ to reproduce observed HI maps. {We show that observations  indicate the need for a high value of $n \gtrsim 80$\,cm$^{-3}$ (although the numerical value is model-dependent), which has important consequences for the dark matter distribution in galaxies.} Our study shows that with innovative methods it is possible to take full advantage of the information content of galaxy images and compare simulations and observations in an interpretable, non--parametric and quantitative manner.
%{\cancel{comparing simulated and real galaxy images in a solid and quantitative way.}}

\end{abstract}

\begin{keywords}
cosmology: theory -- dark matter -- galaxies: formation -- galaxies: kinematics and dynamics -- methods: numerical
\end{keywords}

%%%%%%%%%%%%%%%%%%%%%%%%%%%%%%%%%%%%%%%%%%%%%%%%%%%
\section{Introduction}\label{sec:introduction}
%%%%%%%%%%%%%%%%%%%%%%%%%%%%%%%%%%%%%%%%%%%%%%%%%%%

In the last years hydrodynamic numerical simulations have become one of the most powerful tool to study the formation and evolution of galaxies across cosmic time.
Recent simulations are now able to reproduce a large variety of galaxy properties, {including} (but not limited to) their abundance, stellar masses, rotation velocities, colors, scaling relations, etc. \citep[e.g.][]{Vogelsberger2014,Stinson2013,Dubois2016,Schaye2015,Wang2015,Tremmel2017, Dutton2017,Pillepich2018,Nelson2018,Buck2020b}.

Thanks to the increased computational power, and to many efforts from different groups to improve the modeling of physical processes involved in galaxy formation, several groups have produced extremely high resolution runs of single objects and our local environment  with tens of millions of elements to describe the evolution of dark matter, gas and stars \citep[e.g][]{Aumer2013, Marinacci2014, Hopkins2014,Grand2017, Buck2019,Gutcke2021,Arora2021,Agertz2021}.

Despite these advancements, the multi-scale nature of galaxy formation, spanning from molecular clouds to large scale (cosmological) environment, still precludes the ability to resolve some of the key phenomena that shape galaxies in mass and space, and simulations need to resort to an 'effective description' for some of these processes.
\citep[e.g.][and references therein]{Somerville2015,Vogelsberger2020}.

One possible example of an 'effective description' is the parameterization of star formation in cosmological simulations.
A complete model of star formation would, 
in principle, require to resolve the typical spatial scale of molecular clouds (pc and below) and, at the same time, follow the galactic inflow of gas from cosmological scales of several Mpc, thus bridging some six orders of magnitude in spatial scales. 
In order to overcome the issues related to this challenging dynamical range, cosmological numerical simulations tend to adopt a set of recipes, containing parameters and thresholds, to describe physics below the resolution limit \citep[e.g.][]{Springel2003}. 
For example star formation is usually regulated by a density (temperature) threshold  above (below) which gas particles are eligible to form stars. When it comes to selecting the actual value of this density threshold ($n$), different groups have different approaches. Some groups tend to use values of $n$ around 0.1-1 when measured in particles per cm$^3$ \citep[e.g.][]{Schaye2015, Vogelsberger2014, Nelson2018}, while others prefer to use higher values of $n$, usually in the range $10-100$ \citep[][]{Governato2010,Onorbe2015,Brook2015,Wang2015}.  It is worth noticing that the expected density in the cores of giant molecular clouds should exceed $10^5$ particles per cubic centimeter \citep{McKee2007}, but these values are still out of reach even for the highest resolution simulations of spiral galaxies produced so far \citep[see][for a recent review]{Vogelsberger2020}.
While both approaches (low and high $n$) are similarly successful in reproducing the visible properties of a galaxy, they tend to strongly differ in the predictions for its dark ones. 
For example there are substantial differences in the expected reaction of the dark matter distribution to galaxy formation in low and high $n$ simulations \citep{Benitez2017, Dutton2017}, that might or might not lead us to revisit our current models for the nature of dark matter, as discussed in \cite{Bullock2017}.

The general approach to tune these (somehow) free parameters in the numerical modeling, is to test the final outcome of the simulations, usually galaxies, against observations. In this respect, galaxies scaling relations, as for example, the Tully-Fisher relation \citep{Tully1977}, the size mass relation \citep{Courteau2007}, or the mass metallicity relation \citep{Tremonti2004, Gallazzi2005, Kirby2015}, provide a very good diagnostic tool to select 'good' parameters from 'not-so-good' ones; in the sense that only simulations able to reproduce multiple scaling relations are considered as trustful.

While this approach has its own benefits, it forces one to collapse the large amount of information contained in a simulation into a set of few numbers to be compared with the same few numbers also distilled from complex observations. By this approach a large amount of observational information is neglected as it cannot easily be cast into a single number.
It is then worth asking if it would not be possible to directly compare simulated and real galaxy maps (e.g. stellar or gaseous) in order to use the whole information content of both simulations and observations to help determine which parameters are best suited to capture the physics at work in galaxy formation. 

With the advent of deep learning, the field of machine learning (ML) has matured the tasks of image recognition, classification and segmentation \citep[e.g.][]{Ronneberger2015} greatly benefiting the task of image processing. Such ML techniques have been applied in a large number of astronomical use-cases related to galaxy properties
\citep[e.g.][]{Dieleman2015,Beck2018,Hocking2018,Dawson2020, Buck2021b}.

Unsupervised learning methods have been further used to define kinematic structures of galaxies \citep{Domenech-Moral2012,Obreja2018,Obreja2019,Buck2018,Buck2019b} or to identify accreted stars from disrupted MW satellites in the GALAH survey data \citep{Buder2021}. 

{Relevant in the context of our paper, is the very recent work by \citet{Zanisi2021} comparing small scale morpholigical features of simulated galaxies from the Illustris-TNG project \citep{Pillepich2018} to the SDSS dataset. Those authors find strong disagreement between simulated morphologies and observed galaxies in terms of galaxy morphological features.}

In this paper we aim to use Machine Learning and Artificial Intelligence to {\it quantitatively} compare gaseous (HI) maps from the THINGS
\citep{Walter2008} and VLA-ANGST survey \citep{Ott2012}  with similar maps created from the NIHAO \citep{Wang2015} simulation suite, in order to constrain the numerical value of the density threshold for star formation, $n$, used in simulations.

This paper is organized as follows: in Section \ref{sec:simulations} we present the numerical simulation and their analysis, in Section \ref{sec:obs} and \ref{sec:ML} we describe the observational data and our Machine Learning (ML) algorithm respectively, section \ref{sec:results} is devoted to the presentation of our results which are then discussed in section \ref{sec:conclusions}.

%%%%%%%%%%%%%%%%%%%%%%%%%%%%%%%%%%%%%%%%%%%%%%%%%%%
\section{Simulations} 
\label{sec:simulations}
%%%%%%%%%%%%%%%%%%%%%%%%%%%%%%%%%%%%%%%%%%%%%%%%%%%

The  NIHAO (Numerical Investigation of Hundred Astrophysical Objects) suite of cosmological hydrodynamical simulations \citep{Wang2015,Blank2019}
is based on the  {\sc gasoline2} code \citep{Wadsley2017},
and include Compton cooling and photoionisation and heating from the ultraviolet background following \citet{Haardt2012}, metal cooling, chemical enrichment, star formation and feedback from supernovae and massive stars \citep[the so-called early stellar feedback,][]{Stinson2013}.
The cosmological parameters are set according to the Planck satellites results \citep{Planck2014}. The mass and spatial resolution vary across the whole sample, from a dark matter particle mass of 
$m_{\rm dm}  = 3.4 \times 10^3$ \Msun (and force softening of $\epsilon =100$  pc) for dwarf galaxies to $m_{\sc dm} = 1.4 \times 10^7$ \Msun and $\epsilon = 1.8$ kpc for the most massive galaxies \citep[see][for more details]{Wang2015,Blank2019}.

The NIHAO simulations have been proven to be very successful in reproducing several observed galaxy scaling relations like the Stellar Halo-Mass relation \citep{Wang2015}, the disc gas mass and disc size relation \citep{Maccio2016}, the Tully-Fisher relation \citep{Dutton2017}, and the diversity of galaxy rotation curves \citep{Santos-Santos2018} {as well as the mass-metallicity relation \citep{Buck2021}}.

\subsection{Star Formation}
Star formation is implemented as described in \citet{Stinson2006,Stinson2013}. 
Stars form from cool ($T < 15000$K), dense gas ($\rho > n$) where the density threshold $n$ is measured in particle per cubic centimeters.

In our fiducial NIHAO simulations we adopt
$n = 10$ [cm$^{-3}$] $\sim 50 m_{\rm gas} /\epsilon^3_{\rm gas}$. 
Here, 50 is the number of gas particles used in the SPH smoothing kernel, $m_{\rm gas}$ is the initial mass of gas particles, and $\epsilon_{\rm gas}$ is the gravitational force softening of the gas particle.

We re-run each simulation at two additional star formation thresholds: $n = 0.1$ and $n = 1.0$, without changing any other parameter, and then at a higher density threshold $n=80$ using  half of the softening, in order to keep the same scaling between $n$ and $\epsilon_{\rm gas}$ as in the original NIHAO runs and to ensure, that the resulting gas densities can properly be resolved by the numerics.

\subsection{HI fraction calculation and maps construction}

\begin{figure*}
\includegraphics[width=0.96\textwidth]{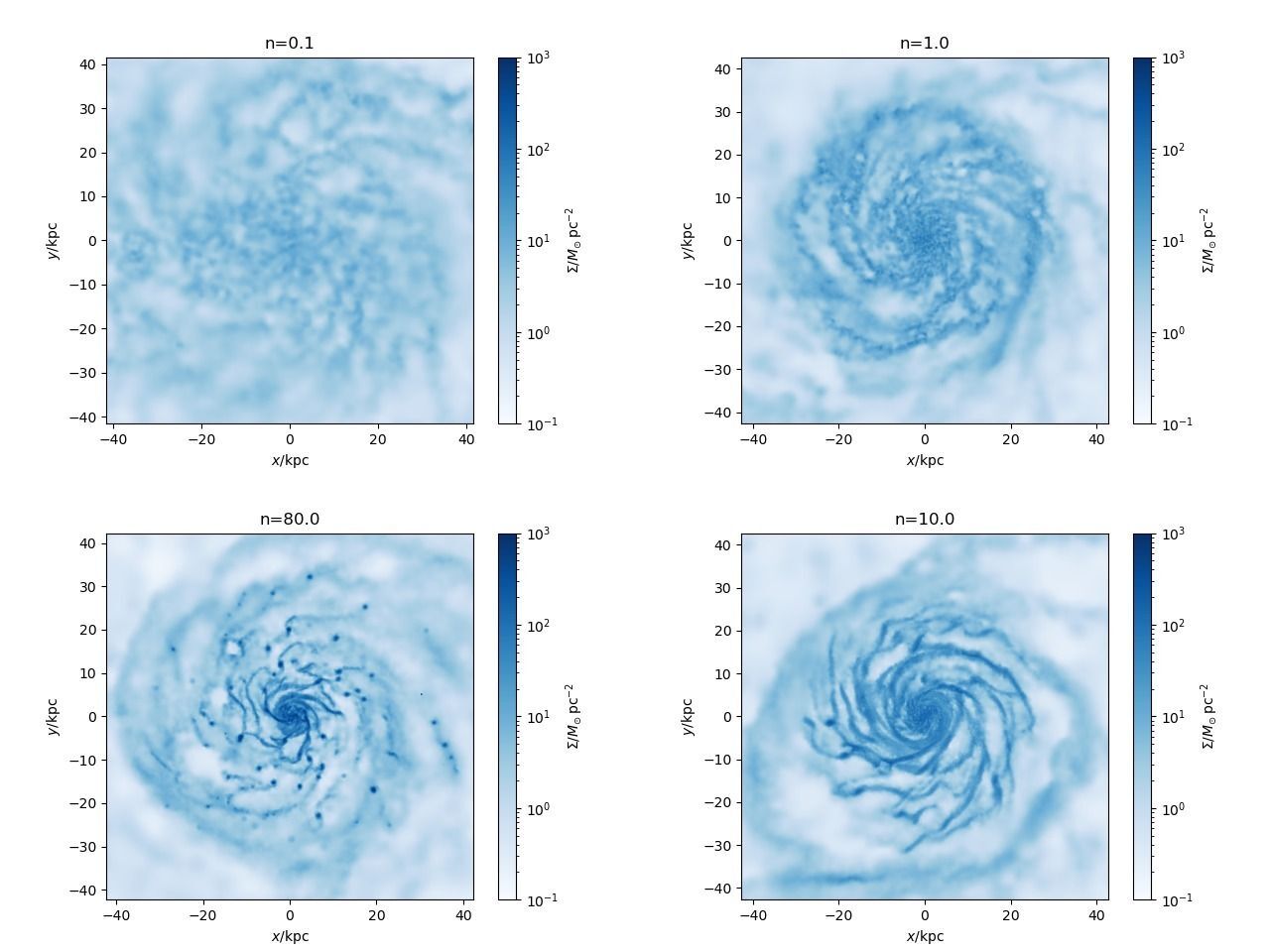}
\vspace{-.25cm}
\caption{HI surface density maps for g8.26e11, a MW-mass like galaxy,  simulated  with different values of the star formation threshold: $n=0.1, 1.0, 10$, and $80$, clockwise from top left.
}
\label{fig:maps}
\end{figure*}

To calculate the neutral hydrogen HI fraction, we followed \citet{Maccio2016} and use the self-shielding approximation described in \citet{Rahmati2013}, based
on full radiative transfer simulations presented in \citet{Pawlik2011}.
The overall effect of this self-shielding approach is to increase the amount of HI 
(relative to the fiducial calculation in {\sc gasoline2}) bringing the simulations in
better agreement with observations \citep{Rahmati2013,Gutcke2016}.
The HI maps have a resolution of 500x500 pixels and have a physical size of 0.4 times the virial radius of the galaxy.
Figure \ref{fig:maps} shows a compilation of HI maps for the same MW-mass like galaxy (face-on) but simulated with different values for $n$.

%%%%%%%%%%%%%%%%%%%%%%
%%%%%%%%%%%%%%%%%%%%%%

\section{Observational Data}
\label{sec:obs}

The observed HI maps have been extracted from The  HI Nearby Galaxy Survey (THINGS)  \citep{Walter2008} and the Very Large Array survey of Advanced Camera for Surveys Nearby Galaxy Survey Treasury galaxies (VLA-ANGST)  \citep{Ott2012}. For this work we used a total of 16 HI maps for galaxies ranging from dwarf irregulars to grand design spirals.

The THINGS maps have pixel sizes of $\sim 20$ to $\sim 100$pc, depending on distance. This is comparable to the simulation rendering for a 500x500 pixel image covering 0.4 of the virial radius of about 200 kpc.
THINGS typically observed the whole HI disk down to densities of a  {few times $10^{19}$\,cm$^{-2}$,}  which usually covers a similar area as 40 percent of the virial radius. We used the natural weighted integrated intensity maps (moment 0) from THINGS and VLA-ANGST. 

For this pilot study we only use a sub-sample of 16 out of the 62 observed maps, as the rest of the galaxies are either too small (or far) or too edge-on for our purposes.
A list of the THINGS and VLA-ANGST galaxies used in this work is given in Appendix \ref{sec:listgalac}.

\section{Machine Learning Models}
\label{sec:ML}

Two different and fully independent deep convolutional neural networks have been used in this work. The first is a Lunar craters detector trained on Lunar surface images, and the second is an image classifier trained on natural world and simulated galaxy images. These two different models have different architectures and were trained on different data sets, thus offering two completely independent views of the problem.

\subsection{Circular structure detector}
\label{sec:ML1}
The first model we deploy is the craters detector presented in \cite{alidib1},
this is a deep convolutional neural network model, based on the semantic segmentation framework %\href{https://github.com/matterport/Mask_RCNN}{\texttt{MaskRCNN}}
\citep{maskrcnnpaper}.

As craters in an image are just quasi-circular brightness anisotropies within a more homogeneous immediate background, we find that this trained model is capable of detecting multi-scale high density ``peaks'' and almost gas-empty ``holes'' in galaxies. To improve the overall model performance and domain transferability, we retrain it on $\sim$ 10$^5$ Lunar, Martian, and Mercurian images for which crater catalogues are present. We moreover employ image augmentation where, in addition to the standard rotations and flipping, we randomly add Gaussian noise, adjust the pixels brightness histogram by a random factor, and add significant tilt angle to the images in an analogue to quasi-edge on galaxies. This turns out to be very important in improving the model's performance, as most of the optical images for example have correlated specific illumination angles that the model might use as features, hindering its transferability. 
{For this model we use the same model as \cite{alidib1}: the \textit{Matterport} implementation of MaskRCNN with the \textit{ResNet101} backbone. The model is trained with Stochastic Gradient Descent as described in \cite{alidib1} and using the same parameters (\texttt{LEARNING$\_$RATE} = $10^{-3}$, 80 epochs, ), \texttt{RPN$\_$ANCHOR$\_$SCALES} = (4, 8, 16, 32, 64), and \texttt{RPN$\_$NMS$\_$THRESHOLD} = 0.7) but on a significantly more complex dataset. It includes images from the Moon's LRO/LOLA global DEM and LRO/LROC WAC Global Optical Mosaic, Mercury's MESSENGER/MDIS Global Mosaic and Global DEM, and finally Mars' MGS/MOLA DEM and Viking Global Mosaic V2. We supplemented the Lunar craters catalogues used by \cite{alidib1} with that of \cite{robbinsMoon}. We used the \cite{robbinsMars} catalogue for Mars, and finally the \cite{mercurycat} catalogue for Mercury.}  

The model is never trained on any galaxy-related images. Inferences on such images is hence purely a form of transfer learning\footnote{Note: A re-training of the network weights on galaxy images is not needed and would further not easily work since MaskRCNN needs ``labeled'' galaxy data. This means for re-training on galaxy images MaskRCNN needs a series of target masks. In any case, at retraining, one would then need to worry about over-fitting, which is not an issue in the use-case explored here.}.  

\subsection{Image classifier}
\label{sec:ML2}
The second model is an image classifier trained using simulated galactic HI maps with the gas density as categorical target.
The model we employed is \texttt{VGG16} \citep{vgg16} that we initiate with the \texttt{imagenet}-pretrained weights. We replace the model's head with fully connected layers: Dense(128, ReLu)-Dropout-Dense(128, ReLu)-Dense(4, softmax).

{ Our model's architecture is shown in Table. \ref{model2table}. We use $L2$ regularization with $\lambda = 0.01$. The model is trained using the \texttt{Adam} optimizer, with a learning rate of $10^{-5}$. As a 5-fold cross-validation scheme is used, the model is trained 5 times on different subset of the data, with 20\% of it kept for the validation score each time. The \texttt{VGG16} layers are initialized with the \texttt{imagenet} weights, but are then allowed to be retrained. \textit{ReLu} is used as activation function everywhere except the output layer, for which we use a \textit{softmax} function.   
\newline

\noindent
\scalebox{0.85}{
\begin{minipage}{\linewidth}
\centering
\captionof{table}{Layer-by-layer description of the classification model, showing the layers' type, output shape, and number of trainable parameters. }
\begin{tabular}{lll}
\hline
Layer (type)                  & Output Shape        & Param \# \\ \hline
vgg16 (Model)                 & (None, 16, 16, 512) & 14,714,688 \\
global\_average\_pooling2d\_1 & (None, 512)         & 0        \\
dense\_1 (Dense)              & (None, 128)         & 65,664    \\
dropout\_1 (Dropout)          & (None, 128)         & 0        \\
dense\_2 (Dense)              & (None, 128)         & 16,512    \\
dense\_3 (Dense)              & (None, 4)           & 516      \\ \hline
\end{tabular}
\label{model2table}

\end{minipage}
}

The four output classes corresponds to four galactic images classes with $n = 0.1, 1, 10, 80$. All of the layers, including \texttt{VGG16}'s, are retrained using $\sim$ 3000 simulated galactic images. Since only $\sim$ 100 simulations are available, we use the same image augmentation routine of Model 1 to considerably increase our data set size. It is worth noting that the training classes are equally distributed and therefore inference should not be affected by minority sampling issues.}

The algorithm performance is evaluated via 5-fold Cross-Validation { and is trained for 10 epochs}. The final trained model reaches 94.6\% categorical accuracy averaged across all folds. This algorithm is never exposed to crater images, and is hence fully independent of model 1 above. The model is finally used in inference mode on the set of observed galaxy images, that it was also never exposed to before. Its output is, for each input image, a size 4 array where each value corresponds to a class probability.

%%%%%%%%%%%%%%%%%%%%%%
\begin{figure*}
\begin{center}
\hspace*{-1cm}                                                      
\includegraphics[scale=0.20]{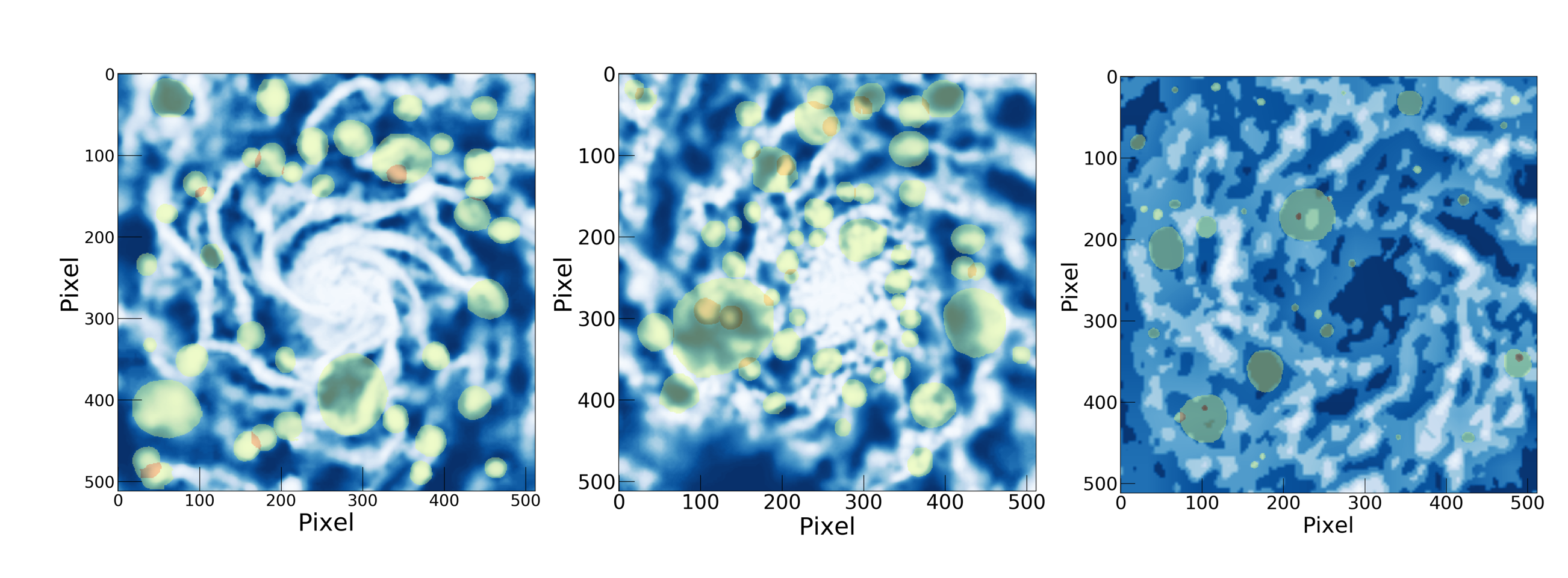}

\caption{Left: Examples from model 1 showing the detected features (green and (when overlapping) orange regions in the image) of simulated galaxies with $n=80$, compared to $n=1$ (center). Notice the higher abundance of smaller scale features close to the galactic core for $n=1$. Right: same as before but for observed galaxy NGC628. Note that in all of the three examples we inverted the colors compared to Fig. \ref{fig:maps} for better visibility: darker region are under-dense with respect to bright ones. The dark center of the observed galaxy is due to a well known large HI hole. 
The total number of detected features is: 40, 95 and 47, from left to right ($n=80$, $n=1$, and observations).}
\label{fig: examplesModel1}
\end{center}
\end{figure*}

\section{Results}\label{sec:results}
\begin{figure}
\begin{center}
\hspace*{-1cm}                                                      
\includegraphics[scale=0.30]{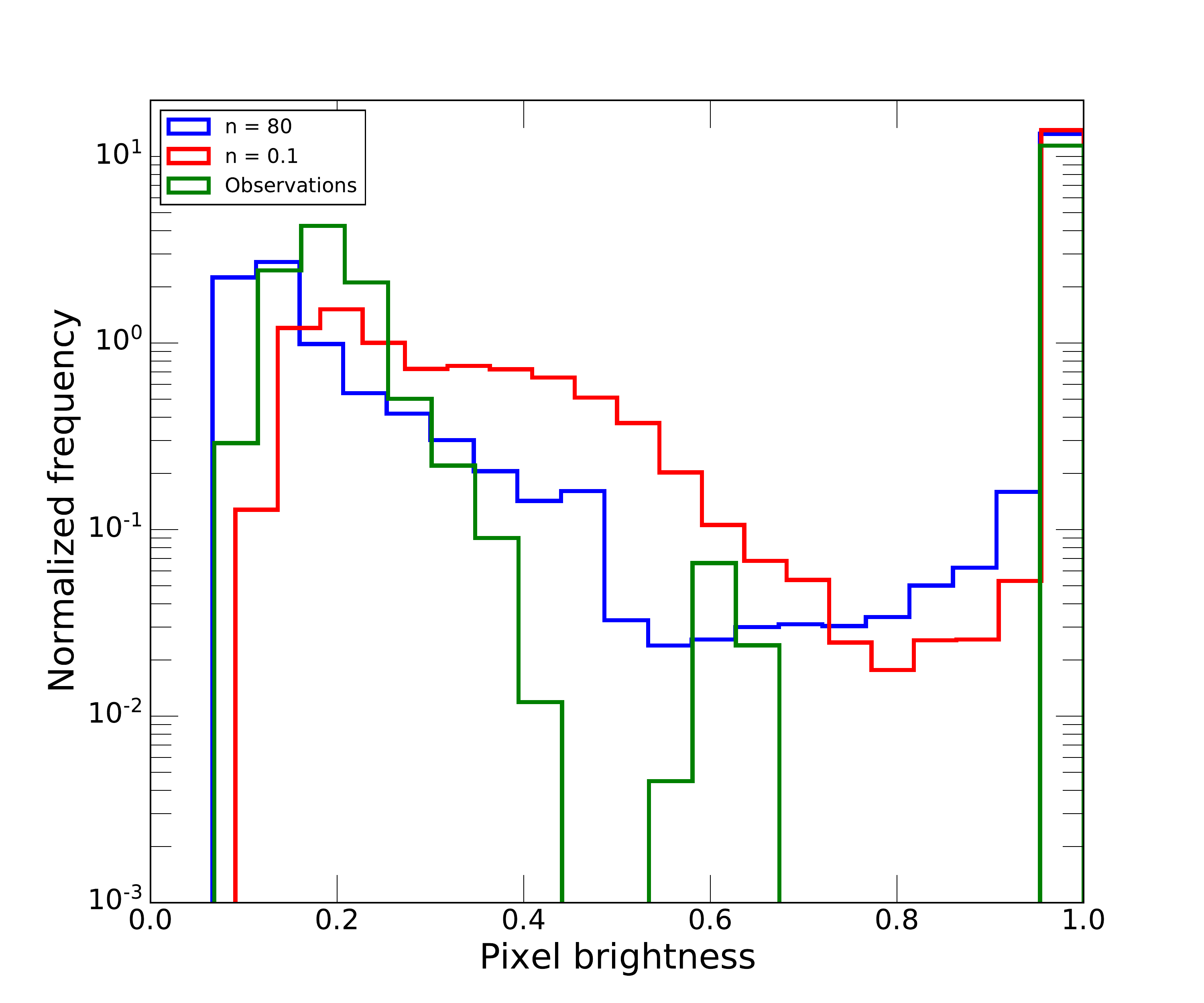}

\caption{Full image pixel brightness distribution averaged for different models and the observed galaxies.}
\label{fig: brighthist}
\end{center}
\end{figure}

In this section we use our two independent neural networks to statistically compare HI maps from simulated galaxies against observed ones, with the aim of constraining whether any specific value for the \textit{effective} threshold density $n$, better reproduces observations. Both  simulated and observed  images are initially read from $x,y$, and flux files, and are logarithmically re-grided and trilinearly interpolated to a given resolution 10$^4$ points in both $x$ and $y$ to generate an image that we then convert to a \texttt{uint8} greyscale image. The image is finally downsampled to 512$\times$512 pixels, the input size the model expects. We note that the raw flux distribution follows a power-law, while the final post-processed flux follow a skewed-Gaussian distribution. Data Gaussianity, while not a strict pre-requisite for deep learning, is usually advantageous in model training. We checked that both simulations and observational images {follow close distributions as can be seen in Fig. \ref{fig: brighthist}. } We finally also note that we performed Fourier analysis on our data, and found that the observational data, and all simulated data, have almost indistinguishable Fourier spectra, highlighting the need of machine learning techniques to compare the two.

\subsection{Features in HI maps}

We run our first ML model (see \ref{sec:ML1}) on the simulated HI maps created
for all galaxies for different values of $n$, the density threshold for star formation. 
Figure \ref{fig: examplesModel1} shows the HI density map for the same galaxy
shown in Figure \ref{fig:maps} simulated with $n=80$ {(left panel)}, superimposed to the
features detected by model 1. Here we used the word {\it feature} to indicate
all the over- and under- dense regions (peaks and craters) identified by the ML algorithm, we do not distinguish the features by size, as this necessitate calibrating the different simulations and observations to a common size scale.
It is worth repeating that this model has never encountered
galactic HI maps before, since it has been trained on planetary maps, yet
it is able to correctly identify the very large majority of the features 
present in the maps.

\begin{figure}
\includegraphics[scale=0.42]{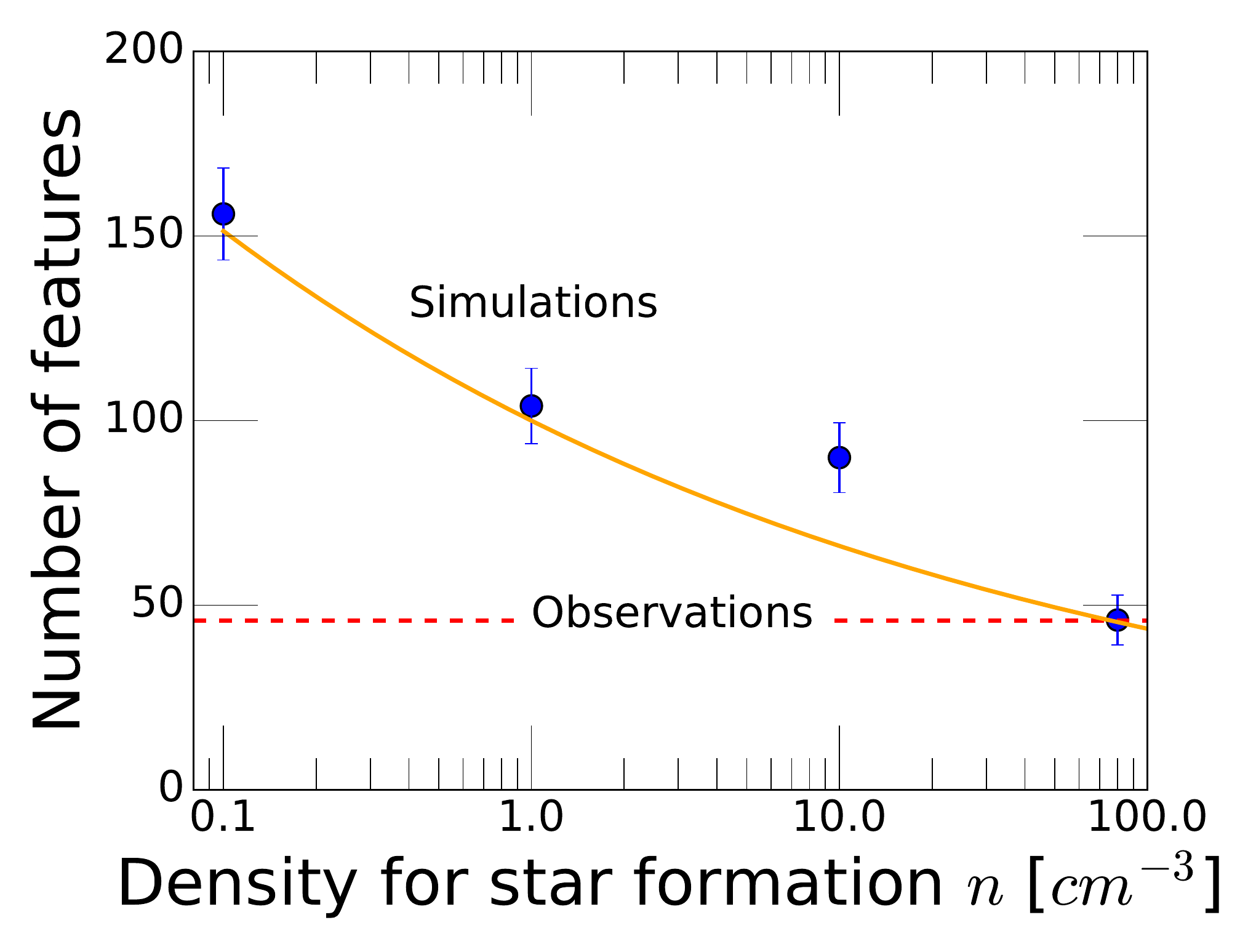}
%\vspace{-.25cm}
\caption{Points with error-bars represent the average number of features per simulated galaxy detected by Model 1, as a function of the galaxies density threshold $n$. The dashed horizontal line is the average number of features recovered in the maps of observed galaxies.}
\label{fig: Model1}
\end{figure}

We then proceed to run our first model on every simulated object to calculate the average number of features detected per galaxy per $n$ density class. The results are shown in  Figure \ref{fig: Model1}, where the average number of features (per galaxy) is plotted as function of $n$. The first interesting result is that there is a clear, negative, 
correlation between  the number of features and the value of the star formation threshold. This can be explained by looking at Figure \ref{fig:maps}: maps with high values of $n$ tend to have fewer major features (e.g. clear spiral arms together with inter-arms under-densities), while runs performed with low values for $n$ have more diffuse maps with a larger number of
small, and uniformly distributed, over-dense and under-dense regions, corresponding to a more flocculent spiral structure. {We note that small scale galactic holes dominate features number for n=0.1, but the numbers for n=80 are too small to make a definitive comparison.}
{More notably, Figure \ref{fig: Model1} shows that it is indeed possible, using  machine transfer learning, to detect image features and separate in } a quantitative way the different $n$-values, going beyond a simple visual inspection of the maps, and hence be employed to quantitatively compare simulations and observations in a holistic way \citep[see also][]{Zanisi2021} which enables to constrain otherwise free model parameters.

%%%%%%%%%%%%%%%%%%%%%%
%%%%%%%%%%%%%%%%%%%%%%

Finally we run the model on the 16 observed HI maps provided by THINGS and VLA-ANGST (that, again, the model have never encountered before), and we get roughly an average of 50 detected features per galaxy (dashed red line in the plot), which is also the same average features number for the simulated galaxies with $n = 80$.

Our results indicate that increasing the threshold density does add observationally constrainable information to the simulated images. We however emphasize that the intersection value $n = 80$ should not be taken at face value, since it is the result of the intrinsic (true) star formation threshold \citep[$\sim 10^4$ particle per cubic centimeter][]{McKee2007} convolved with the resolution of both the observations and the simulations. However, it clearly shows that if the numerical resolution is high enough, a higher star formation threshold (i.e. a multi-phase ISM) is observationally preferred. We conclude that there is a strong need for cosmological simulations to improve their current models in order to provide realistic galaxy models that can be trusted. We also note that the observed image resolutions (or equivalently redshift) might affect the value of $n$ at which the two curves intersect, but this does not affect our central conclusion that $n$ significantly affects the number of features in simulated images, and thus the two curves have to meet for some value of $n$.

Similarly, there might be secondary variations with other galaxy properties such as stellar mass or gas mass, but since we did not train our model on galaxies (but on planetary craters!) this 
should not (substantially) affect our conclusions, 
and we expect our model to be transferable to different resolutions,
masses and redshifts. Nevertheless, in future work, it might be interesting to explore this in more detail. Finally, Since no ground truth for the number of density features exist, it is not possible to evaluate the performance of this model quantitatively. For this reason, we use another model below in order to strengthen our conclusions.

\subsection{Galaxy classification according to $n$}

We confirm the results of Model 1 using the independent Model 2. This model was trained to classify  simulated galaxies images by their $n$ value. 
We now use it to classify the observed galaxies, objects that the model was never exposed to before. 

\begin{figure}
\includegraphics[scale=0.42]{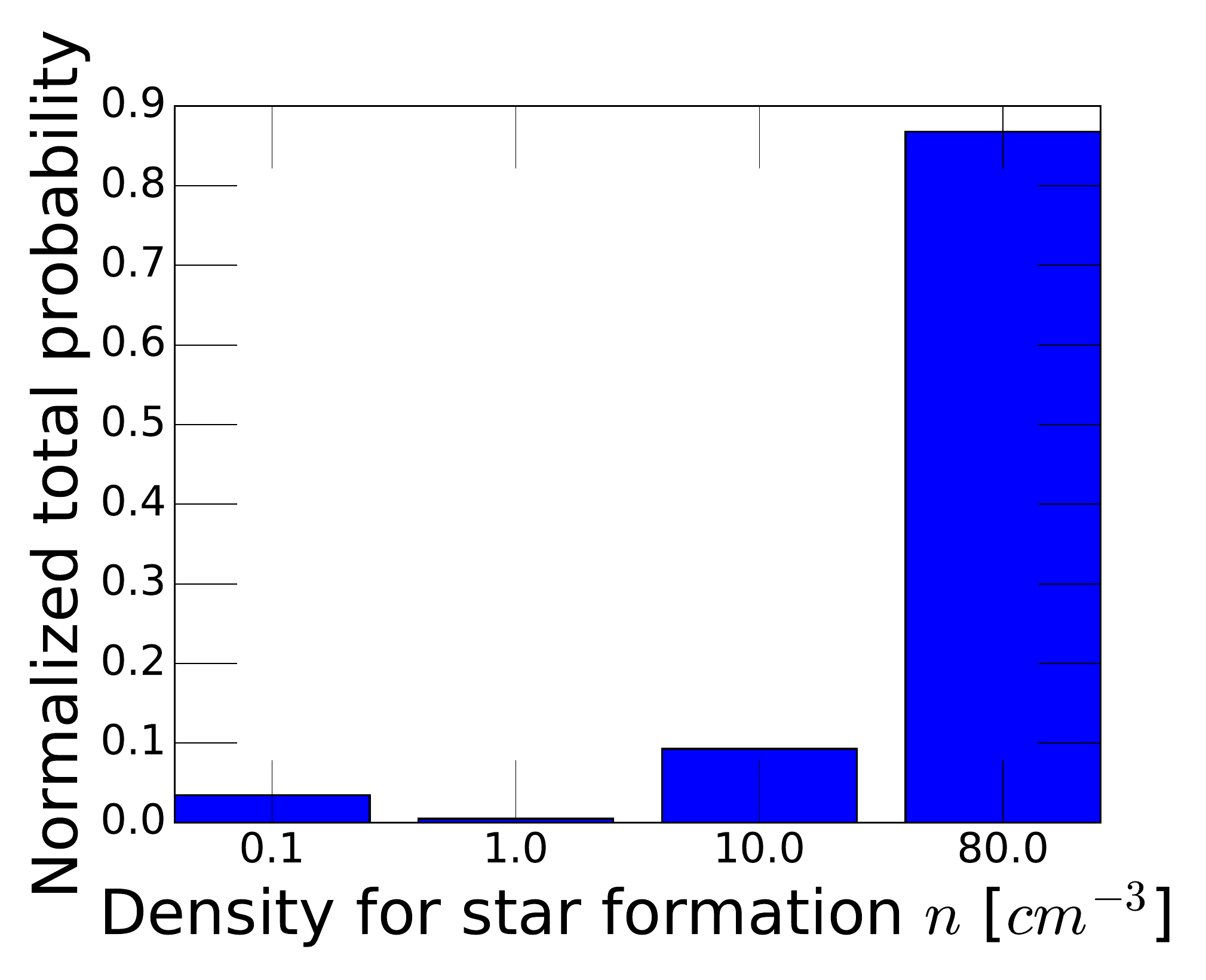}
%\vspace{-.25cm}
\caption{The probability of maps from observed galaxies to belong to the different $n$ classes. There is a clear preference for $n=80$.}
\label{fig: Model2}
\end{figure}

In Figure \ref{fig: Model2}, we show the probability belonging to a given $n$ class for the observed galaxies, based on their HI maps. 
 The model infers with very high probability ($\ge$ 93\%) a value of $n=80$ for 13 out of the 16 HI observed images, while it classifies two also as having $n=80$ but with certainty $\sim 66\%$. Just one galaxy is classified as $n=10$. 
When considered all together, the vast majority (larger than 85 percent) of the observed HI maps are found by the model to be in the $n = 80$ class, with a very high degree of algorithmic confidence,
especially considering that the training set was fully balanced and hence no particular class was favoured (see section \ref{sec:ML2}).

{We however emphasize that these probabilities are are not posteriors, but simply model confidence through the softmax output. We therefore calculate the ROC-AUC score, and find its value to be $\sim$ 0.9. This implies that the model can distinguish accurately between the different class points.  }

{As a further robustness check, we change the model to cast the problem as a binary classification by merging the n=0.1 and n=1; and n=10 and n=80 classes \citep{galacn1,galacn2}. We found this to increase the neural net's preference for higher $n$ values, with all of the observed galaxies now classified as n=10/80. }

{These results are consistent with the conclusions from Model 1, and strengthen its methodology and findings, especially that} the two models {employ very different algorithms and} have nothing in common and were trained on very different sets of data. 
Moreover the two different models try to answer two very different questions: Model1 is the answer to {\it which simulations statistically match obervations?}, while Model2 is the answer to {\it If the observations were simulated, which simulation would they be?}, underlying the complementary views of the problem provided by the two models.

%%%%%%%%%%%%%%%%%%%%%%
%%%%%%%%%%%%%%%%%%%%%%

\section{Discussion and Conclusions}
\label{sec:conclusions}

Numerical simulations of galaxy formation are an incredible powerful tool to study the complex network of effects and relations that leads to the creation and evolution of a galaxy in a universe dominated by dark matter and dark energy.
Despite the impressive advancements in the last years  \citep[e.g][]{Pillepich2018, Buck2020,Agertz2020}, simulations still rely 
on a set of parameters that need to be calibrated against observations.

In this paper we present a first attempt to directly use images (i.e. 2D maps) of simulated and real galaxies in order to constrain simulations parameters, without the need to 'extract' from such images galaxy parameters like mass, size, luminosity etc.

In our pilot study we used two completely independent Machine Learning algorithms applied to HI maps of real and simulated galaxies to constrain the value of $n$, the density threshold for star formation, using a set of values commonly employed in numerical simulations \citep[][]{Dutton2019}.

The first ML model  has been trained to identify craters in Lunar and planetary maps, and it is here employed to find and count over-dense gas peaks and low density gas ``holes'' in the HI maps (generally referred as features). Simulated galaxies, from the NIHAO project, show a strong, negative correlation between the number of features and the value of $n$, allowing to quantify the differences between the different models. When applied to real, observed maps, the ML algorithm  detects a number of feature very close to the one of $n=80$ (particle per cubic centimeter) model, confirming the need of large values of $n$ to correctly reproduce observations \citep{Buck2019b}. Due to the absence of an unbiased performance metric for the first model, we use another method in order to strengthen our results.

The second model, independent from the first, is an image classifier, that we trained to predict the value of $n$ for the HI maps.
When this model is applied to the observed maps, that the model has never encountered before, it infers with very high probability ($\ge$ 86\%) a value of $n=80$ for all real  HI images.

These two models, taken together, not only reinforces the need to use a large ($n \ge 10$) value for
the density threshold, but, more importantly, shows that it is possible to take advantage of the large amount of information contained in real and simulated galaxy maps to constrain numerical simulations, without the need to reduce them to a limited set of global parameters like stellar mass, total mass, color etc.

It is worth stressing that while the parameter used in this study, $n$, is a numerical parameter and not a physical one, the approach tested here can also be applied to models that circumvent the need for a density threshold and tie the star formation model to the local H$_2$ fraction or the virial parameter of the gas. In that sense the approach presented here gives a universal, unbiased and automated way of constraining/calibrating/differentiating between valid galaxy formation models (not only limited to the star formation process).

Given that large amount of current and incoming galaxy surveys, that will realise stunning new images of nearby and distant galaxies \citep[e.g.][]{Emsellem2021}, the approach presented in this work, has the potential to bring new insights on how to correctly parametrise complex physical processes in numerical simulations.

\section*{Data availability}

The data underlying this article will be shared on reasonable request to the corresponding author.
\section*{Acknowledgements}

This material is based upon work supported by Tamkeen under the NYU Abu Dhabi Research Institute grant CAP$^3$.
The authors  gratefully acknowledge the Gauss Centre for Supercomputing e.V. (www.gauss-centre.eu) for funding this project by providing computing time on the GCS Supercomputer SuperMUC at Leibniz
Supercomputing Centre (www.lrz.de) and the High Performance Computing resources at New York University Abu Dhabi.



\bibliographystyle{mnras}
\bibliography{ref} % if your bibtex file is called example.bib

\appendix

\section{List of THINGS and VLA-ANGST galaxies used in this work}\label{sec:listgalac}

NGC3184, NGC3521, NGC3621, NGC404, NGC4214, HoII, IC2574, NGC5055, NGC5194, NGC5236, NGC5457, NGC628, NGC6946, NGC2403, NGC925, NGC2903.

\label{lastpage}
\end{document}